\documentclass[iop,tighten,apjl]{emulateapj}
\usepackage{amsmath}     
\usepackage{wasysym}           
\usepackage{graphicx}
\usepackage{amssymb}
\usepackage{epstopdf}
\usepackage{mathrsfs}
\usepackage{natbib}
\usepackage{physymb}
\usepackage{color}

\begin{document}
\title{Variability in tidal disruption events: gravitationally unstable streams}
\shorttitle{Variability in tidal disruption events}
\author{Eric R. Coughlin\altaffilmark{1} and Chris Nixon\altaffilmark{2}}
\shortauthors{\sc{Coughlin \& Nixon}} 
\affil{JILA, University of Colorado and NIST, UCB 440, Boulder, CO 80309}
\email{eric.coughlin@colorado.edu, chris.nixon@jila.colorado.edu}
\altaffiltext{1}{Department of Astrophysical and Planetary Sciences, University of Colorado, UCB 391, Boulder, CO 80309}
\altaffiltext{2}{Einstein Fellow}

\begin{abstract}
{We present simulations of the tidal disruption of a solar mass star by a $10^6M_{\astrosun}$ black hole. These, for the first time, cover the full time evolution of the tidal disruption event, starting well before the initial encounter and continuing until more than 90\% of the bound material has returned to the vicinity of the hole.} Our results are compared to the analytical prediction for the rate at which tidally-stripped gas falls back. We find that, for our chosen parameters, the overall scaling of the fallback rate, $\dot{M}_{\rm{fb}}$, closely follows the canonical $t^{-5/3}$ power-law. However, our simulations also show that the self-gravity of the tidal stream, which dominates the tidal gravity of the hole at large distances, causes some of the debris to recollapse into bound fragments before returning to the hole. {This causes $\dot{M}_{\rm{fb}}$ to vary significantly around the $t^{-5/3}$ average.} We discuss the implications of our findings in the context of the event \emph{Swift} J1644+57.
\end{abstract}


\section{Introduction}
When a star comes within a supermassive black hole's (SMBH) tidal radius $r_{\rm t} \simeq R_*(M_h/M_*)^{1/3}$, where $R_*$ and $M_*$ are the stellar radius and mass, respectively, and $M_{\rm h}$ is the black hole mass, the tidal force exerted by the hole across the star is sufficient to overcome its self-gravity, resulting in its destruction. Early studies of these tidal disruption events (TDEs) demonstrated that roughly half of the disrupted stellar debris is bound to the black hole \citep{lac82, ree88}, meaning that it will eventually return to the tidal radius, dissipate {energy} through shocks \citep{eva89, koc94, gui14b}, and form an accretion disk. The resulting accretion power is then capable of producing a highly luminous event, and many have already been detected \citep{kom99, gez08, bur11, cen12, bog14}.

Considerations of the star at the time of disruption show that the rate at which the bound, tidally-stripped material returns to the hole decreases as $t^{-5/3}$ \citep{phi89}, and early simulations supported this estimate \citep{eva89}. More recently, authors investigated the consequences of the stellar composition on the rate of return \citep{lod09}, {}{illustrating that the early fallback stages have a more complex temporal behavior that depend on the density stratification.} \citet{gui13} showed that the {}{pericenter distance} can have dramatic effects on the rate of return of debris. General relativistic Lense-Thirring \citep{sto12} and apsidal \citep{bon15} precession and recompression shocks \citep{gui14b} have also been investigated.

Most of the recent work on TDEs has been performed with the aid of numerical simulations. For example, \citet{gui14b} used a grid code (see also \citealt{shi15}) to follow the evolution of the tidally disrupted debris from the initial encounter with the black hole to its eventual return to pericenter, while \citet{bon15} used a smoothed particle hydrodynamics (SPH) code to achieve the same feat (see also \citealt{hay15}). In these cases, however, the set of parameters used to model the tidal disruption was somewhat unphysical -- \citet{gui14b} used a $10^3M_{\astrosun}$ black hole when following the return of the debris to the pericenter radius, while \citet{bon15} used a more tightly bound stellar progenitor ($e = 0.95$ for their most parabolic simulation, $e$ being the stellar eccentricity). Both of these choices were made to reduce the computational cost at the expense of reality. {}{Instead, a more likely mass for an SMBH is $10^5$ -- $10^8M_{\astrosun}$, and the infalling star is initially so far from the hole that its orbit is effectively parabolic ($e\simeq1$).}

A full TDE with realistic parameters has not yet been modeled because of the extreme set of spatial and temporal scales. In spite of these difficulties, {}{we present the first simulation that has resolved the full duration  of a TDE with a solar progenitor on a parabolic orbit, {}{the periapsis of which coincides with the tidal radius,} and a $10^6M_{\astrosun}$ SMBH. The simulation runs from well before the initial tidal interaction to long after the most bound debris returns to pericenter, corresponding to roughly ten years after disruption. This enables us to calculate explicitly the rate of return of tidally stripped debris to the black hole, demonstrating the effects of self-gravity which dominate that of the hole when the material is near apoapsis. In section 2 we provide analytic formulae for the shape and density distribution of the stream as it recedes from the hole, and {we develop arguments which suggest the dominance of self-gravity over the tidal field of the hole at large radii}. In section 3 we describe {the initial conditions and numerical method used to simulate the encounter}. Section 4 presents the results of the simulations and their analysis. We conclude and discuss further implications of our findings in section 5.
\newline
\newline

\section{Self-gravity}
The forces acting on the debris are self-gravity, the tidal shear across the width of the stream, and pressure. If we consider a small volume within the stream of length $\delta{R}$ and mass $\delta{M}$, then the self-gravity of the stream dominates the tidal field if

\begin{equation}
\rho_{0} \gtrsim \frac{3}{2\pi}\frac{M_h}{r^3} \label{sgdeltaM},
\end{equation}
where $\rho_0 = 3\delta{M}/(4\pi\delta{R}^3)$.

We can calculate when this condition is satisfied by approximating the stream as a circular cylinder of width $H$. In this case, the differential amount of mass $dM$ contained in the stream is

\begin{equation}
dM = \pi{H}^2\rho\,dx \label{dMdx},
\end{equation}
where $x$ is the length measured along its center. If the specific energy distribution of the gas is frozen in at disruption, then the curve traced out by $x$ is a function solely of the position of a given gas parcel when the star is at pericenter, which we will denote by $R_p = \mu{R_*}$, with $-1\le\mu\le1$. Modeling the star as a polytrope \citep{han04}, the differential amount of mass contained in $d\mu$ is \citep{cou14}

\begin{equation}
\frac{dM}{d\mu} = \frac{1}{2}M_*\xi_1\frac{\int_{\mu\xi_1}^{\xi_1}\Theta(\xi)^{n}\xi{}\,d\xi}{\int_0^{\xi_1}\Theta(\xi)^{n}\xi^2d\xi} \label{dMdeta},
\end{equation}
where $n$ is the polytropic index of the gas, $\Theta(\xi)$ is the solution to the Lane-Emden equation and $\xi_1$ is the first root of $\Theta(\xi)$.

We can parametrize $x$ by the functions $r(\mu)$ and $\theta(\mu)$, where $r$ and $\theta$ are solutions to the equations of motion of a test particle in the Newtonian gravity of the hole:

\begin{equation}
r = \frac{\ell^2}{GM_h}\frac{1}{1+e\cos\theta} \label{roftheta},
\end{equation}
\begin{equation}
r^2\dot\theta = \ell \label{thdot}.
\end{equation}
Here $\ell$ and $e$ are the specific angular momentum and eccentricity, respectively, of the gas parcel with position $\mu$, and are given by

\begin{equation}
\ell = \sqrt{2GM_hr_t}\big{(}1-\mu\,{q}^{-1/3}\big{)} \label{ellofeta},
\end{equation}
\begin{equation}
e = 1-2\mu\,q^{-1/3} \label{eofeta},
\end{equation}
where $q \equiv M_h/M_*$. We have assumed here that the gas parcels initially move with the center of mass of the star and $\beta = r_t/r_p = 1$. By numerical integration of equation \eqref{thdot} we can determine $\theta(\mu)$, and $r(\mu)$ is then found by using that result in equation \eqref{roftheta}.

The infinitesimal distance along the curve is $dx = d\mu\sqrt{(r')^2+r^2(\theta')^2}$, where primes now denote differentiation with respect to $\mu$. Using this result and equation \eqref{dMdeta} in equation \eqref{dMdx} gives

\begin{equation}
\rho = \frac{M_*\xi_1}{2\pi{}H^2\sqrt{(r')^2+r^2(\theta')^2}}\frac{\int_{\mu\xi_1}^{\xi_1}\Theta(\xi)^{n}\xi{}d\xi}{\int_0^{\xi_1}\Theta(\xi)^{n}\xi^2d\xi} \label{rhostream}.
\end{equation}
Using this equation in condition \eqref{sgdeltaM} gives the range in $\mu$ over which self-gravity dominates the tidal shear. 

The pressure of the material can also resist self-gravity. However, even if a small overdensity of radius $\delta{R}$ and mass $\delta{M}$ is pressure-supported, it must satisfy the Jeans condition $\delta{R} \lesssim c_s\sqrt{\pi/(\rho\,{G})}$. If $\delta{R}$ exceeds this value, any perturbation will cause material to clump around $\delta{M}$.

Finally, the local velocity gradient can potentially stabilize the stream. The divergence of the velocity decreases the density within the stream on a timescale $\tau_{div} \simeq 1/(\nabla\cdot{v})$. If $\tau_{div} \gtrsim \tau_{ff}$, where $\tau_{ff} \simeq \delta{R}^{3/2}/\sqrt{G\delta{M}}$ is the local free-fall time, self-gravity will overcome the stabilizing effect of the divergence of the flow.

By balancing the tidal shear and self-gravity in the transverse direction, \citet{koc94} found that $H \propto r^{1/4}$ for a $\gamma = 5/3$ equation of state. We will use this when considering how the density (equation \ref{rhostream}) varies with $r$ (see Figure \ref{fig:rhoplotsim}).

\section{Simulation setup}
{To test our analytical reasoning above and the canonical $t^{-5/3}$ fallback rate, we now employ three dimensional hydrodynamic simulations.} We use the SPH code {\sc phantom} \citep{pri10, lod10} to simulate the tidal encounter. {\sc phantom} was designed for running high resolution hydrodynamic simulations with complex geometries \citep{nix12a, nix12b, mar14a, mar14b, nea15}, and is therefore well suited to studying TDEs.

For our simulations we model the black hole as a Newtonian sink particle situated at the origin. {Self-gravity of the stellar material is modeled using a k-D tree alongside an opening angle criterion, calculating directly the forces arising over short distances \citep{bar86}. The opening angle for the simulations presented here was 0.5 (we ran an additional simulation with a more accurate opening angle of 0.3 and found negligible differences with the run presented here). The gas retains a polytropic equation of state, i.e., only adiabatic energy gains and losses are included (we also performed simulations with shock heating and, with our parameters, found negligible differences; see also \citealt{lod09}).

We initialized our runs by placing $10^6$ particles in the configuration of a $\gamma = 5/3$ polytrope; this is achieved by first placing the particles on a close-packed sphere. This configuration is then stretched to achieve the correct polytropic density distribution, the resulting polytrope then placed at a distance of $10\,r_t$ from the hole with its center of mass on a parabolic orbit, {}{its location at periapsis equal to the tidal radius} (equations \eqref{roftheta} and \eqref{thdot} with $\ell = \sqrt{2GM_hr_t}$ and $e = 1$). The distance of $10\,r_t$ allows the initial configuration time to relax before interacting with the black hole.

Here we are primarily interested in exploring the effects of self-gravity on the stream evolution, and so we include self-gravity at all stages of the simulation. We are also interested in comparing the theoretical rate of return of bound material (not necessarily equal to the accretion rate onto the black hole), $\dot{M}_{\rm fb} \propto t^{-5/3}$, with that inferred from the simulation. Therefore, once the disrupted debris is beyond periapsis, the accretion radius of the black hole, initially well inside the tidal radius, is extended to $r_{\rm acc} \simeq 3\,r_t \, (\simeq 120 R_{\rm g})$. Once the bound material returns to the point of disruption, it is removed from the simulation.

\section{Results}

\begin{figure*}[htbp] 
   \centering
   \includegraphics[width=.495\textwidth, height=.43\textwidth]{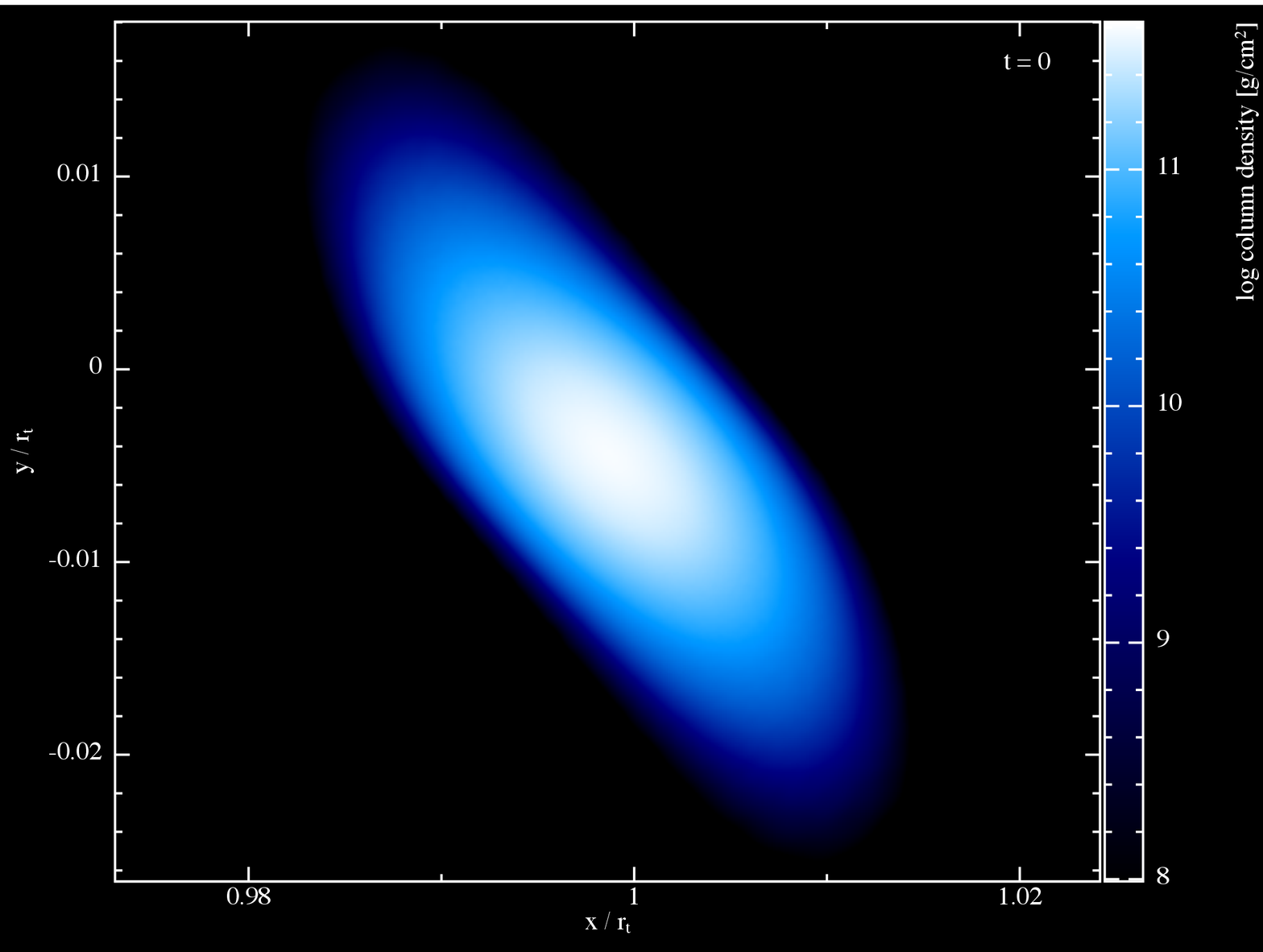} 
   \includegraphics[width=.495\textwidth, height=.43\textwidth]{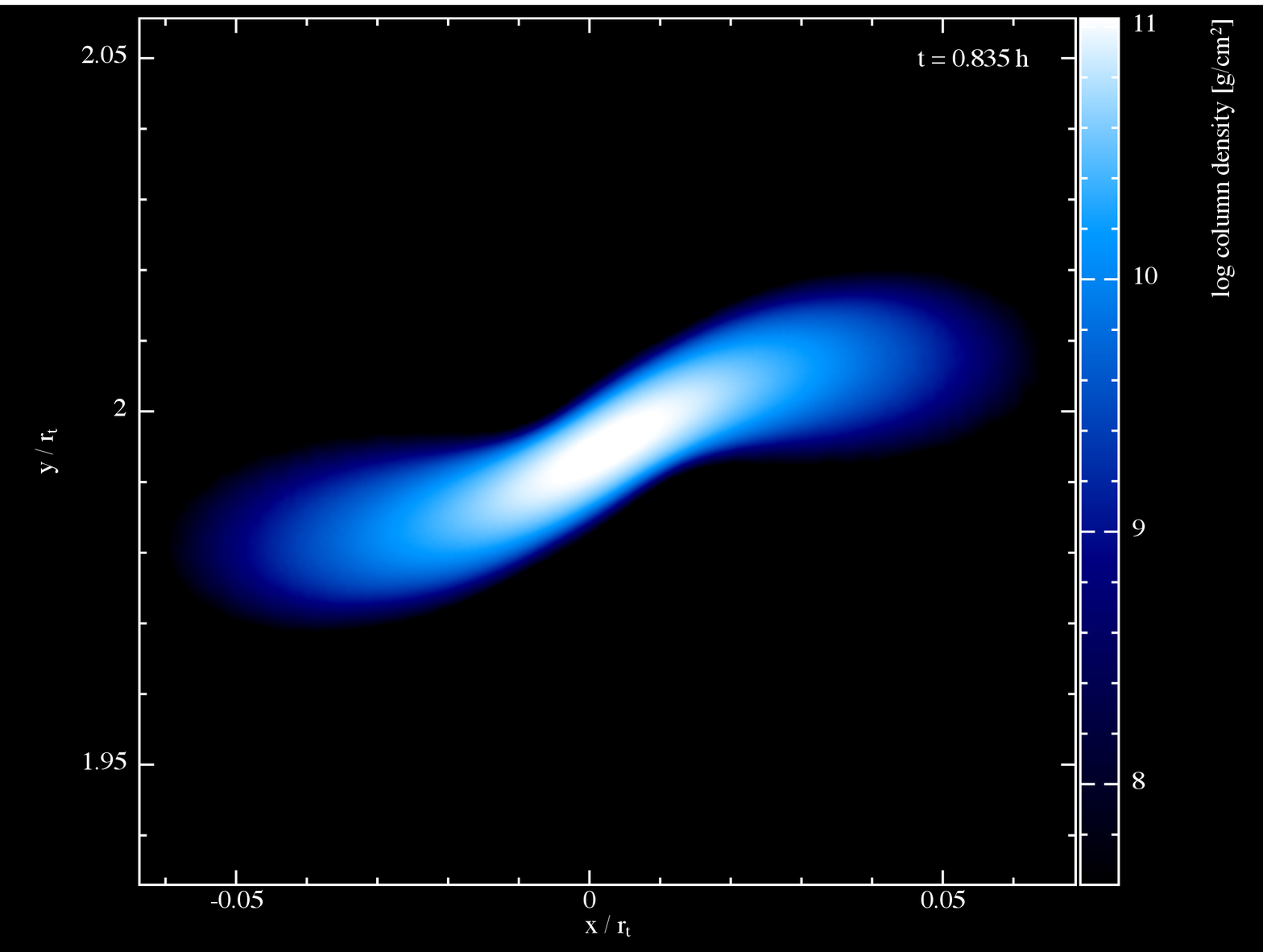}
   \includegraphics[width=.495\textwidth, height=.43\textwidth]{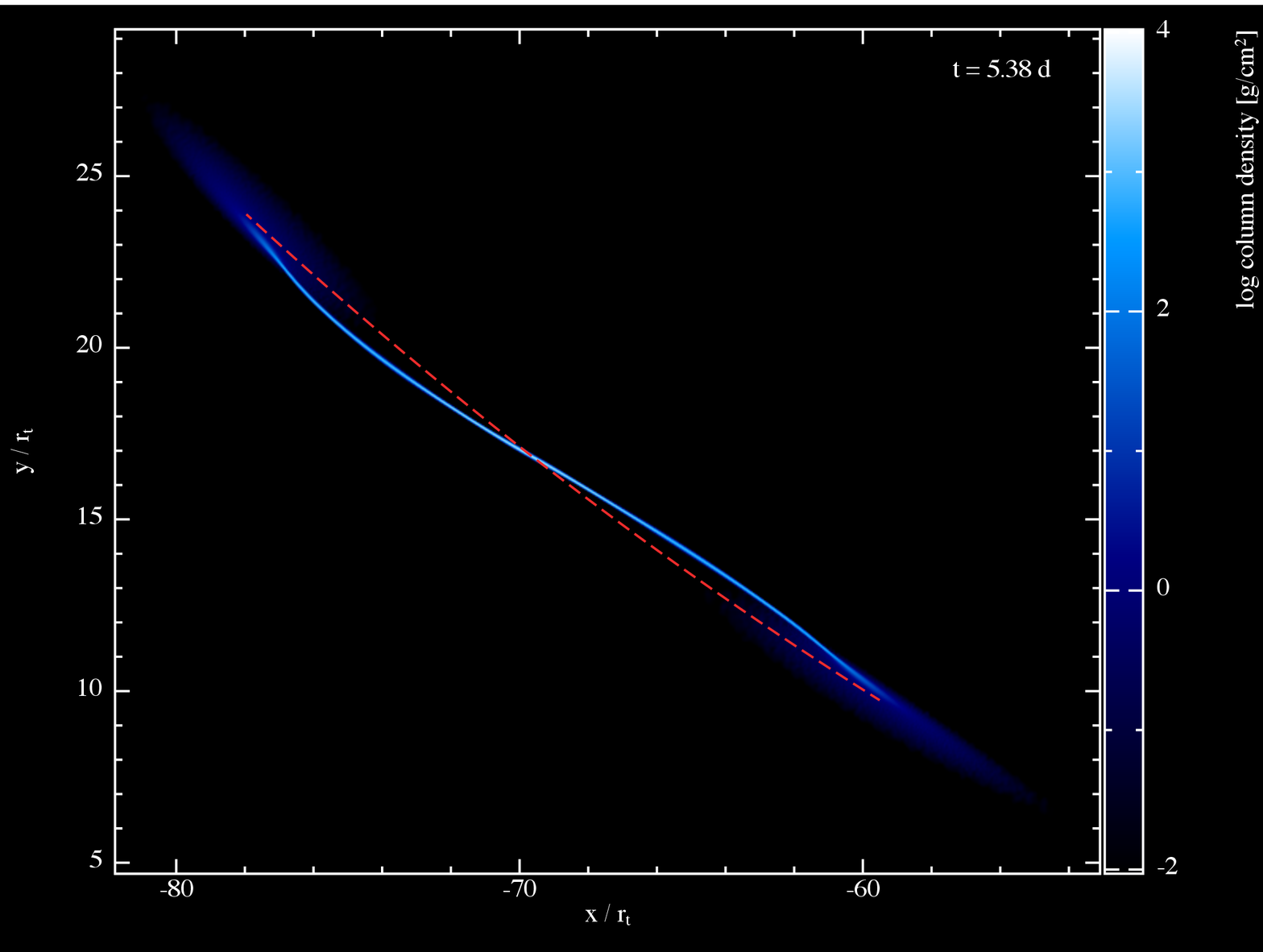} 
   \includegraphics[width=.495\textwidth, height=.43\textwidth]{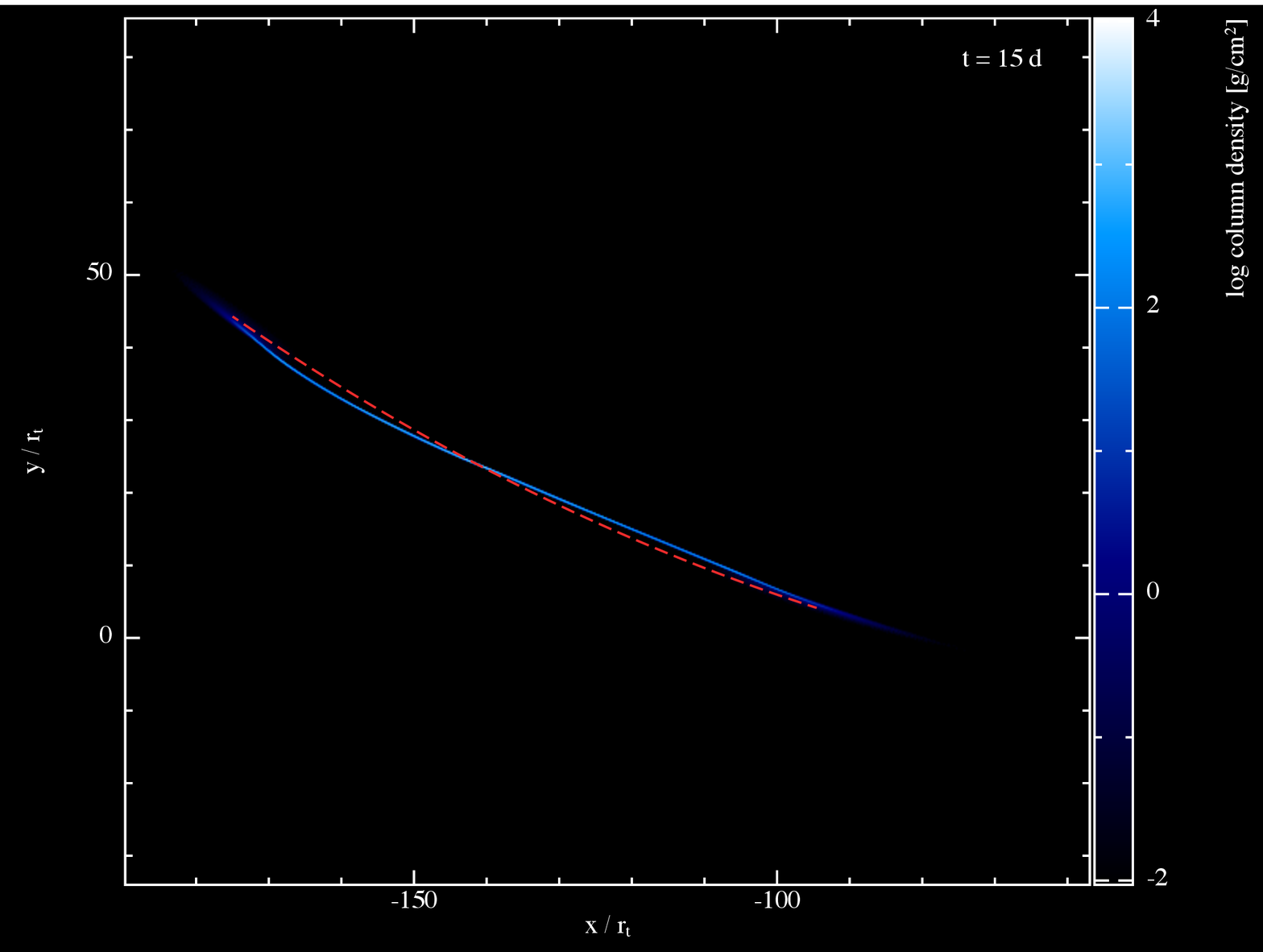} 
   \includegraphics[width=.495\textwidth, height=.43\textwidth]{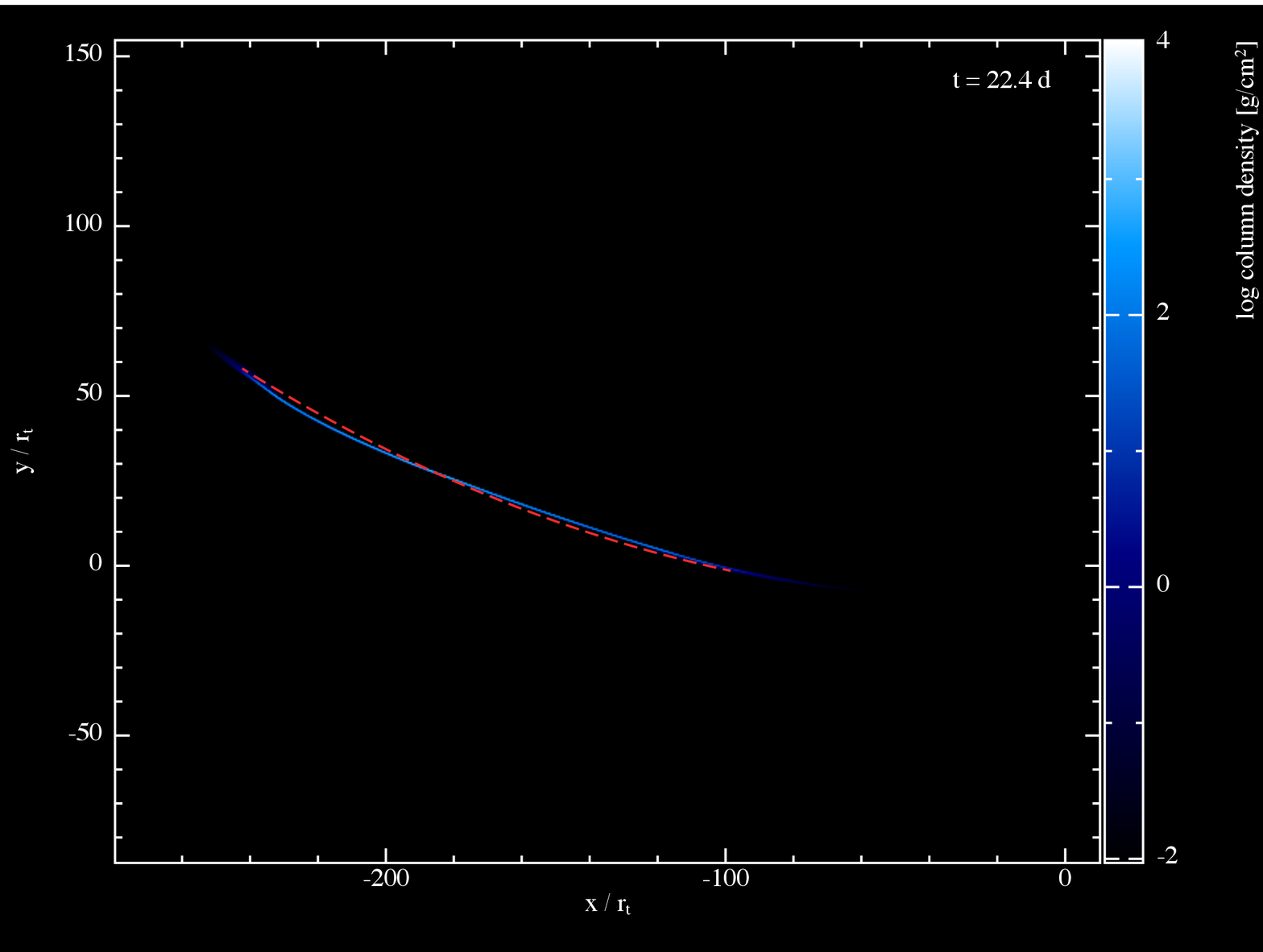} 
   \includegraphics[width=.495\textwidth, height=.43\textwidth]{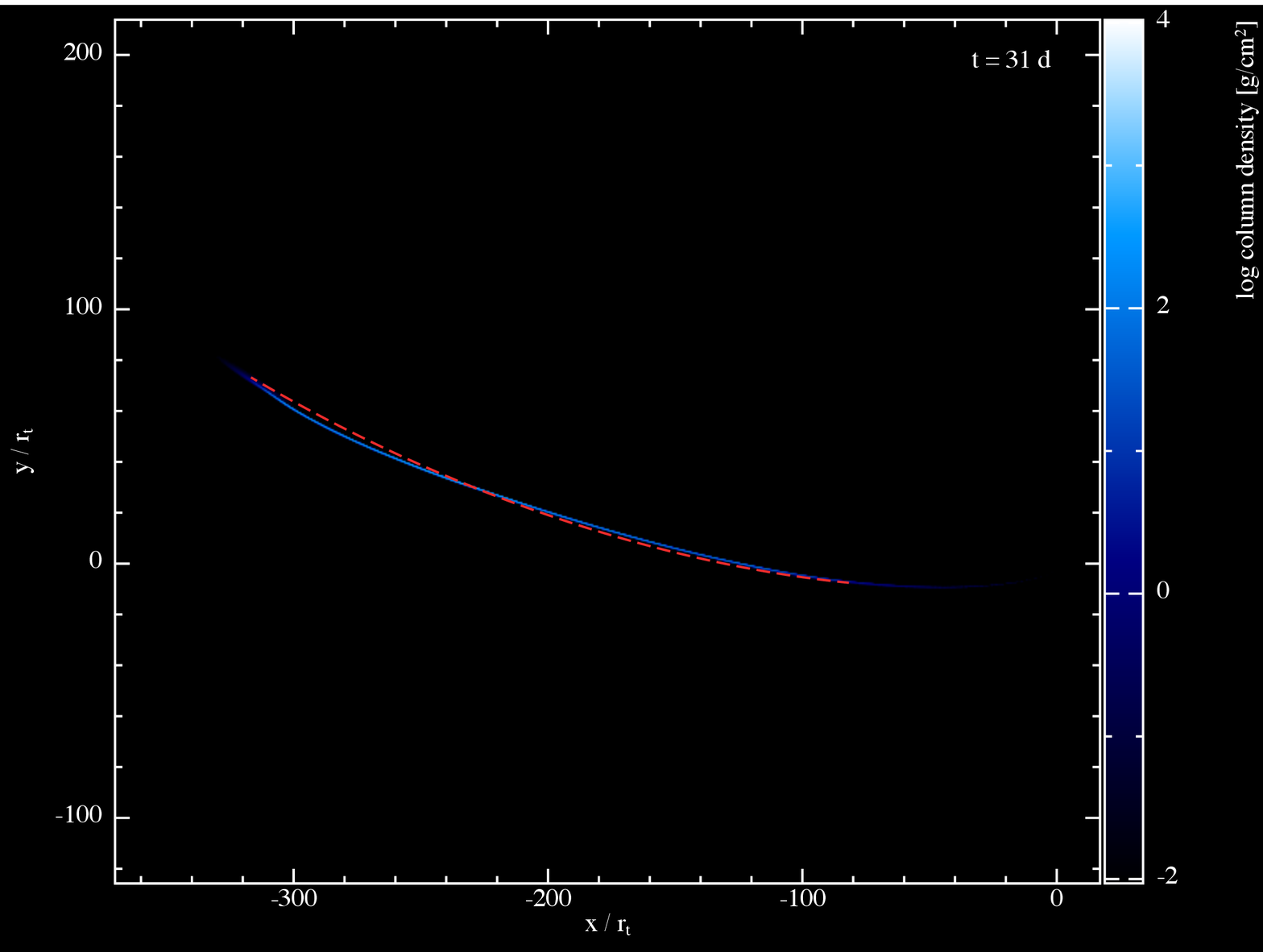} 
   \caption{The star at the time of disruption (top, left), where the colors indicate the column density and distances are measured in units of tidal radii ($\simeq 7\times10^{12}$ cm $=100R_{\astrosun}$), $0.835$ hours after periapsis (top, right; the top two figures coincide with the bottom-left panels of Figure 4 a) and b), respectively, of \citealt{lod09}.), and 5.38 days (middle, left), 15 days (middle, right), 22.4 days (bottom, left) and 1 month (bottom, right) after disruption. The red, dashed curves on the bottom two rows indicate the analytic predictions. }
   \label{fig:streamplot}
\end{figure*}

\begin{figure}[htbp] 
   \centering
   \includegraphics[width=3.3in]{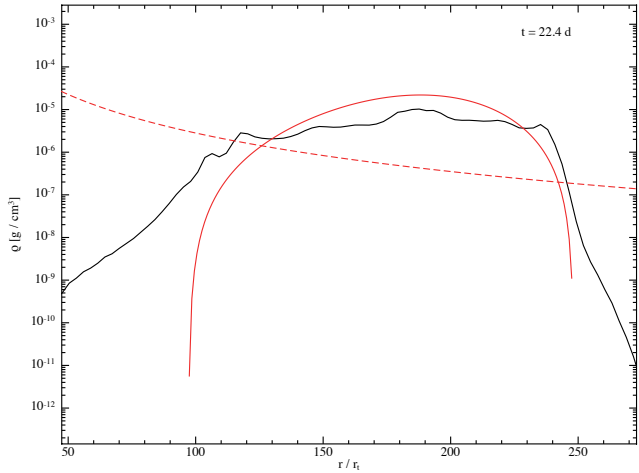} 
   \caption{The computed density of the stream (solid, black curve), the analytic prediction (red, solid curve; equation \ref{rhostream}), and the density at which self-gravity becomes important (red, dashed curve; equation \ref{sgdeltaM}) as functions of $r$ (spherical distance from the hole) at 22.4 days from the point of disruption. Points above the red, dashed curve are self-gravitating. The numerical solution extends to smaller radii because the star at the time of disruption is not spherical (see also \citealt{lod09}).}
   \label{fig:rhoplotsim}
\end{figure}

\begin{figure}[htbp] 
   \centering
   \includegraphics[width=3.3in]{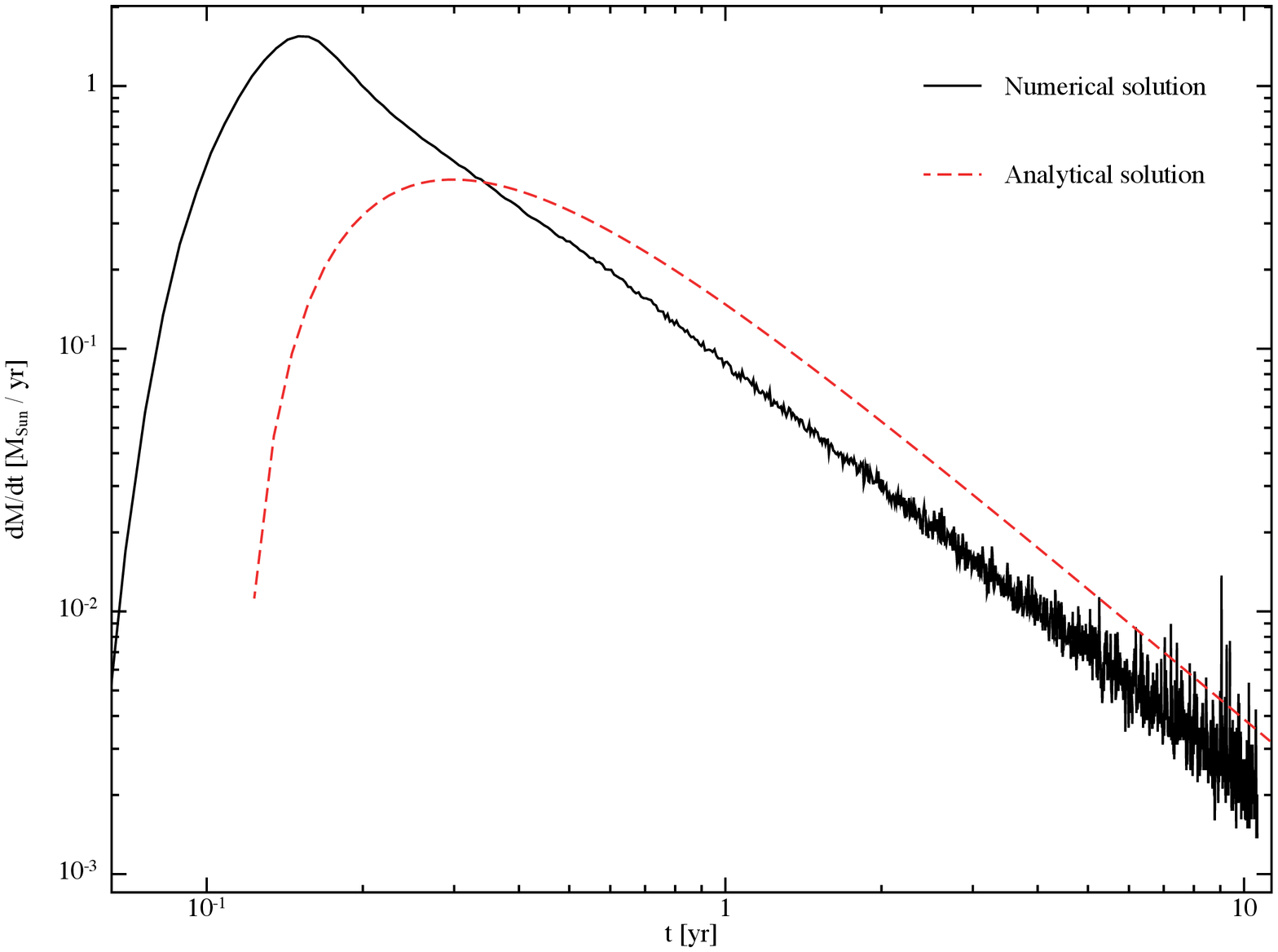} 
   \caption{The fallback rate in solar masses per year calculated from the simulation (black, solid curve) and the analytic prediction (red, dashed curve) as functions of time in years. The temporal shift between the numerical solution and the analytic prediction is from the stellar distortion at the time of disruption (see text for further details). The variability at late times is due to the accretion of clumps formed in the self-gravitating stream. The variability starts about 2 years after disruption; however, this timescale depends upon the details of the TDE and its environment, and therefore may be shorter.}
   \label{fig:mdotsim}
\end{figure}

The top, left panel of Figure \ref{fig:streamplot} shows the density of the debris stream at periapsis, while the top, right is 0.84 hours after disruption; these reproduce Figure 4 of \citet{lod09}. The bottom four show, from left to right and top to bottom, 5.38 days, 15 days, 22.4 days and one month after disruption. The red curves indicate the analytic predictions; the fact that they provide good fits to the data means that the particles approximately trace out Keplerian orbits in the potential of the black hole. 

The black points in Figure \ref{fig:rhoplotsim} show the density distribution along the stream at $t = 22.4$ days after disruption; the solid, red curve gives the analytic prediction (equation \eqref{rhostream} with ${H = R_{\astrosun}(r/r_t)^{1/4}}$) and the dashed, red curve is the density at which the self-gravity of the debris and the tidal field of the hole are equal along the stream (the right-hand side of equation \eqref{sgdeltaM}; self-gravity dominates for points above this curve). The wings present in the simulated values, which were also found by other authors \citep{lod09}, are due to the fact that the shape of the star at the time of disruption is not spherical (see the left-most panel of Figure \ref{fig:streamplot}). 

Figure \ref{fig:mdotsim} shows the rate of return of material to pericenter, $\dot{M}_{fb}(t)$, from the simulation (solid, black curve) and the analytic solution (red, dashed curve). The analytic estimate was determined by using equation \eqref{dMdeta} and the fact that $\mu(t) = (t/T)^{-2/3}$, where $T = 2\pi{M_h}/(M_*\sqrt{GM_h})(R_*/2)^{3/2}$ (\citealt{cou14}). The time at which material returns to pericenter is slightly earlier than that predicted analytically, a consequence of the tidal distortion at the time of disruption (see Figure \ref{fig:mdotsim}). The asymptotic scaling $\dot{M}_{fb} \propto t^{-5/3}$ is closely followed by the numerical solution at late times, reaffirming that the gas parcels follow approximately Keplerian motion. In our simulation, half of the bound material is accreted by about one year -- shortly after the peak in the fallback rate. 

\begin{figure}[htbp] 
   \centering
    \includegraphics[width=3.3in]{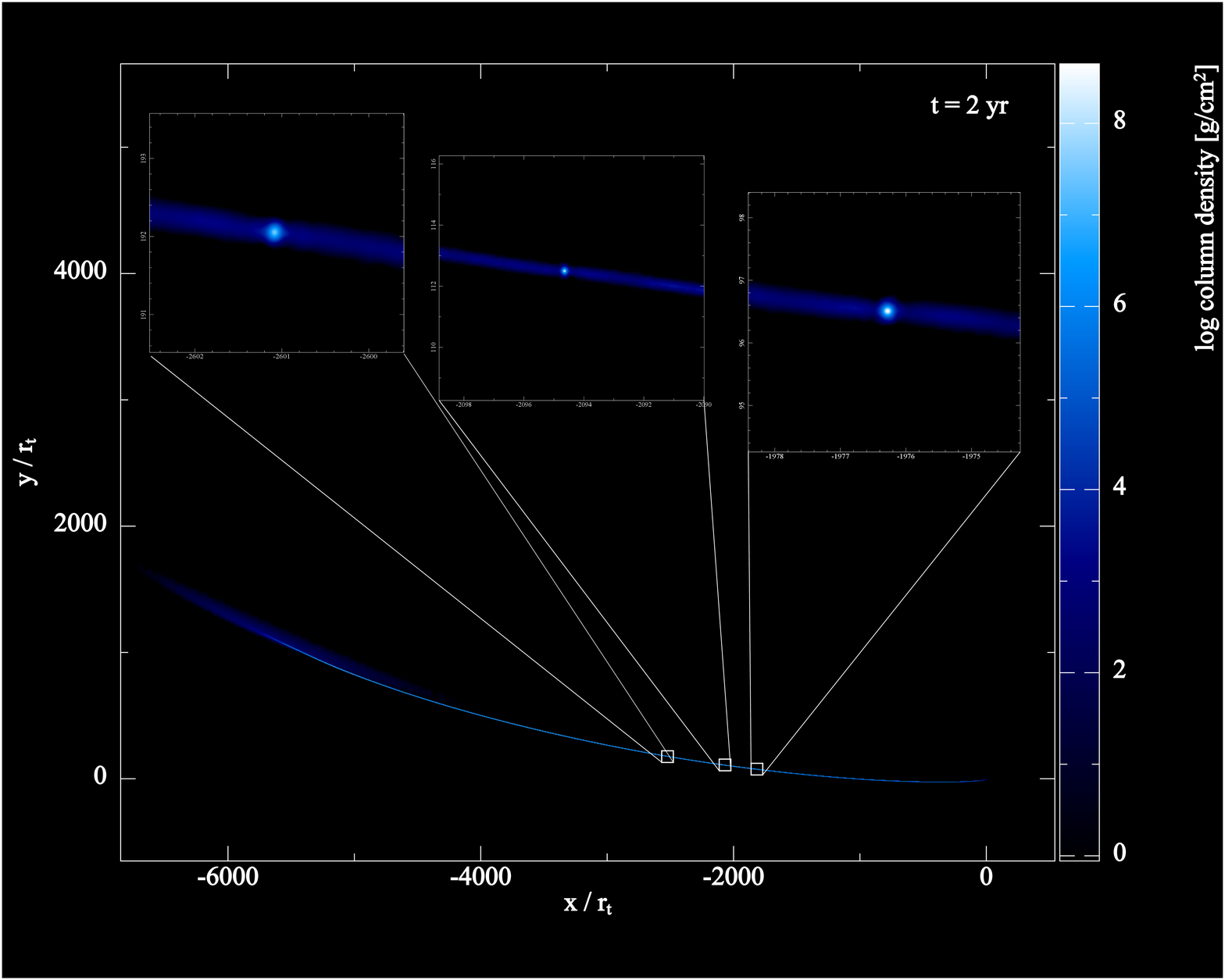}
   \caption{The density two years after disruption and, in the insets, {}{close-ups of the} fragments that forms along the stream.}
   \label{fig:streamclump}
\end{figure}
Figure \ref{fig:mdotsim} also shows that the fallback rate quasi-periodically deviates from the {}{$t^{-5/3}$ law} at late times. This scatter is due to the fact that the stream fragments, forming small gravitationally bound clumps that enter the accretion radius at discrete times, and is \emph{not} simply numerical noise. To emphasize this point, Figure \ref{fig:streamclump} shows the stream two years after disruption and insets that focus on {}{clumps} near the center of the stream.

\section{Discussion and conclusions}
We have presented the results of a TDE in which a solar mass star (with its pericenter at the tidal radius and its center of mass on a parabolic orbit) was disrupted by a $10^6M_{\astrosun}$ black hole. Contrary to past investigations, we resolved the full duration of the TDE -- from the initial encounter between the undisturbed star and the hole to long after the most bound, tidally-stripped debris has returned to periapsis. When finished, more than 90\% of the bound material was accreted. {Crucially, we are the first to follow this long-term evolution for a solar mass star on a parabolic orbit around an SMBH.}

The two main aims of this paper are (1) to determine the effects of self-gravity on the stream far from the black hole (see Figures \ref{fig:mdotsim} and \ref{fig:streamclump}), and (2) to compare the rate of return of bound material to the analytic estimates (Figure \ref{fig:mdotsim}). The fiducial rate $\dot{M}_{fb} \propto t^{-5/3}$ was determined by assuming that the specific energy distribution of the material was frozen in at the time of disruption. However, the self-gravity of the material, which is important at large distances from the hole (Section 2), has the potential to alter the specific energy distribution (see also \citealt{gui13}). It is therefore necessary to resolve the full disruption process with self-gravity included at every step to determine the true rate of return. 

Our simulations demonstrate that the fallback rate of debris closely mimics the theoretically-predicted one, the biggest discrepancy arising from the return time of the most tightly bound material. Therefore, the material, in agreement with intuition, follows approximately Keplerian orbits about the hole. Interestingly, however, we also found that the fallback rate at late times tends to over and under-estimate the {$t^{-5/3}$ law}, a feature that is due to the fragmentation of the stream into gravitationally-bound clumps. When one of these clumps is accreted, the fallback rate spikes above its average value, while the rate is reduced when the lower density debris between clumps returns to periapsis. 

We have shown analytically and numerically (see equation \eqref{sgdeltaM} and Figure \ref{fig:rhoplotsim}) that the local self-gravity of the stream dominates the tidal field of the hole at large radii.  We also find that the pressure within the sphere {}{of} the {}{center} inset of Figure \ref{fig:streamclump} is $p\simeq 10^3\text{ dyn}/\text{cm}^2$. Conversely, the pressure necessary for maintaining hydrostatic equilibrium is $p_{eq} \simeq GM\rho/R \simeq 4\pi{G}\rho^2R^2/3$, where the radius of the clump in Figure \ref{fig:streamclump} is $R \simeq 0.3\,r_t$ ($= 30R_{\astrosun}$) and the density is $\rho\simeq 10^{-6}\text{ g}/\text{cm}^3$, which gives $p_{eq} \simeq 10^6\text{ dyn}/\text{cm}^2$. Finally, the computed divergence of the velocity just outside the clump gives a divergence timescale of $\tau_{div} \simeq 1/(\nabla\cdot{v}) \simeq 10^{7}$ s, while the infall timescale is $\tau_{ff} \simeq 1/\sqrt{G\rho} \simeq 4\times10^6$ s. {}{This shows that gas pressure and the local shear of the stream are also incapable of overcoming self-gravity.} 

The tidal field of the hole, gas pressure, and the local velocity shear are all incapable of supporting the clump in Figure \ref{fig:streamclump} against its own self-gravity. This finding suggests that the original stream was gravitationally unstable, and small perturbations resulted in its fragmentation. For the results presented here, these perturbations are induced by the discreteness of the numerical method. To substantiate this claim, we ran the simulation with $10^4$  and $10^5$ particles and found that clumps formed sooner with decreasing particle number. However, our pressure smoothing length, equal to the gravitational smoothing length, was always at least as small as the Jeans radius $R_J = c_s\sqrt{\pi/({G}\rho)}$ {}{(see Figure \ref{fig:honRJ_rho})}; the collapse was therefore always resolved \citep{tru97, bat97}. {}{Additionally, the first clumps in the simulation presented here form around a month after disruption; at this time, the smoothing length at the center of the stream is roughly 0.1 times the width of the stream, meaning that the forces are clearly resolved at this time {}{(see also Figure 5, which shows that the collapse itself is also resolved)}. Finally, we note that our SPH simulations are not susceptible to artificial fragmentation \citep{hub06}, and our results demonstrate that any physical perturbation imposed on a stream that satisfies \eqref{sgdeltaM} is unstable to collapse.}

\begin{figure}[htbp] 
   \centering
   \includegraphics[width=3.3in]{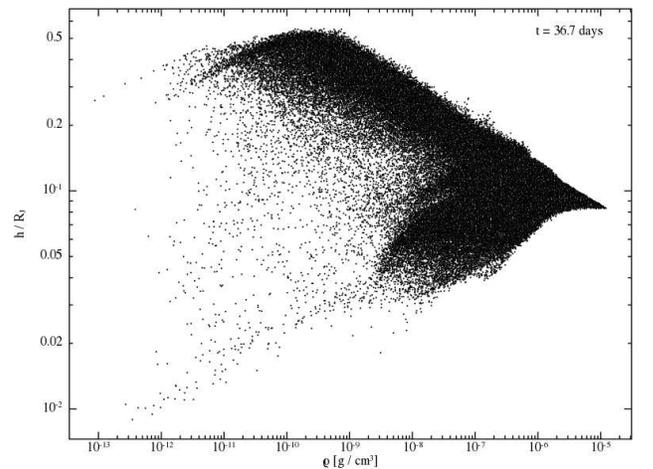} 
   \caption{{}{The ratio $h/R_J$, $h$ being the smoothing length and $R_J$ the Jeans radius, as a function of density at a time when the first clumps start to form.}}
   \label{fig:honRJ_rho}
\end{figure}

To further investigate the instability of the stream, we performed additional runs with $10^5$ and $10^6$ particles and seeded the initial distribution of debris with a small, but resolved, perturbation. In these cases, clumps formed at almost the exact same time with nearly identical masses. Therefore, a resolved noise field imposed on top of the otherwise smooth distribution of stellar debris results in converged fragmentation. This finding further supports the fact that the stream is gravitationally unstable, and any small physical perturbation to the distribution of the debris will cause it to fragment. We will present a more complete analytical and numerical analysis of the gravitational stability of the stream in a future publication. 

Real, fragmentation-inducing perturbations could be caused by a number of physical processes. For example, the interaction of the debris with any material surrounding the hole, the density distribution of which is neither necessarily smooth nor homogeneous, could cause local deformities within the stream. In light of this, the variability in the light curve of the event \emph{Swift} J1644+57 \citep{bur11, lev11, zau11}, the recently-observed, jetted TDE, could be interpreted as abrupt changes to the accretion rate induced by the fallback of bound clumps. This notion is supported by the lack of any ultraviolet and optical emission from the event, which is indicative of a large amount of dust present in the circumnuclear environment \citep{bur11}; interactions between the stream and this natal environment could have initiated early fragmentation of the stream, causing variability in the fallback rate on a timescale commensurate with observations.

Finally, we note that {}{while} the mass contained in the fragments in Figure \ref{fig:streamclump} is $\delta{M} \simeq 0.005M_{\astrosun}${}{, there is a range of clump masses. Therefore, we suggest that the object G2 -- the clump of material observed near the galactic center \citep{bur12} -- and other such clouds, could have been produced by the tidal disruption of a star in the recent past.} \citet{gui14a} reached a similar conclusion; their clump, however, was formed by fluid instabilities generated through the interaction of a stream of debris with an ambient medium. Here, on the other hand, self-gravity causes the stream to fragment.

\acknowledgments ERC was supported in part by NASA Astrophysics Theory Program grant NNX14AB37G, NSF grant AST-1411879, and NASA's Fermi Guest Investigator Program. CN was supported for this work by NASA through the Einstein Fellowship Program, grant PF2-130098. Both authors thank Phil Armitage, Mitch Begelman, Andrew King, and Jim Pringle for useful comments, and we thank the Boulder, CO parks and recreation committee for their upkeep of the Valmont disk golf course, which has been a site of most fruitful discussion. {We used {\sc splash} \citep{pri07} for the visualization.}  This work utilized the Complexity HPC cluster at the University of Leicester which is part of the DiRAC2 national facility, jointly funded by STFC and the Large Facilities Capital Fund of BIS.

\end{document}